\documentclass[journal]{IEEEtran}
\newcommand{\subparagraph}{}
\usepackage{amsmath,amsfonts,amssymb}
\usepackage{graphicx}
\usepackage[colorlinks=true, allcolors=blue]{hyperref}
\def\UrlBreaks{\do\/\do-}
\usepackage{lipsum}
\usepackage{caption}

\DeclareMathOperator*{\argmin}{arg\,min}
\usepackage{cite}
\usepackage{titlesec}
\usepackage{flushend}
\usepackage{algorithm}
\usepackage[super]{nth}
\usepackage[noend]{algpseudocode}
\usepackage{ctable}
\definecolor{green1}{RGB}{6, 155, 0}
\definecolor{blue1}{RGB}{0, 0, 210}
\usepackage{url}
\makeatletter
\g@addto@macro{\UrlBreaks}{\UrlOrds}
\makeatother

\begin{document}
\title{Generating Anthropomorphic Phantoms Using Fully Unsupervised Deformable Image Registration with Convolutional Neural Networks}


\author{\IEEEauthorblockN{Junyu Chen,
Ye Li,
Yong Du,
Eric C. Frey}\\ 
\IEEEauthorblockA{Russell H. Morgan Department of Radiology and Radiological Science, Johns Hopkins Medical Institutes}\\
\IEEEauthorblockA{Department of Electrical and Computer Engineering, Johns Hopkins University}\\ Baltimore, MD, USA}


%
\maketitle
\begin{abstract}
\textit{Objectives:} Computerized phantoms play an essential role in various applications of medical imaging research. Although the existing computerized phantoms can model anatomical variations through organ and phantom scaling, this does not provide a way to fully reproduce anatomical variations seen in humans. However, having a population of phantoms that models the variations in patient anatomy and, in nuclear medicine, uptake realization is essential for comprehensive validation and training. In this work, we present a novel image registration method for creating highly anatomically detailed anthropomorphic phantoms from a single digital phantom. \textit{Methods:} We propose a deep-learning-based registration algorithm to predict deformation parameters for warping an XCAT phantom to a patient CT scan. This proposed algorithm optimizes a novel SSIM-based objective function for a given image pair independently of the training data and thus is truly and fully unsupervised. We evaluate the proposed method on a publicly available low-dose CT dataset from TCIA. \textit{Results:} The performance of the proposed model was compared with that of several state-of-the-art methods, and outperformed them by more than $8\%$, measured by the SSIM and less than $30\%$, by the MSE. \textit{Conclusion:} A deep-learning-based unsupervised registration method was developed to create anthropomorphic phantoms while providing "gold-standard" anatomies that can be used as the basis for modeling organ properties. \textit{Significance:} Experimental results demonstrate the effectiveness of the proposed method. The resulting anthropomorphic phantom is highly realistic. Combined with realistic simulations of the image formation process, the generated phantoms could serve in many applications of medical imaging research.
\end{abstract}

\section{Introduction}


\IEEEPARstart{C}{omputerized} phantoms for nuclear medicine imaging research have been built based on  anatomical and physiological models of human beings. They have played a crucial part in evaluation and optimization of medical image reconstruction, processing and analysis methods\cite{Christoffersen2013, Zhang2017, Chen2019, Abdoli2013}. Since the exact structural and physiological properties of the phantom are known, they can serve as a gold standard for the evaluation and optimization process. The 4D extended cardiac-torso (XCAT) phantom \cite{Segars2010} was developed based on anatomical images from the Visible Human Project data. This realistic phantom includes parameterized models for anatomy, which allows the generation of a series of phantoms with different anatomical variations. These phantoms have been used in Nuclear Medicine imaging and CT research \cite{He2008, Ghaly2016, Li2018, Nakada2015, LeeKappler2017, LeeKapplerPolster2017, Kidoh2017}, as well as in various applications of deep learning\cite{Gong2018, Gong2019, Lee2017}. 

In the XCAT phantom, changing the values of parameters that control organ anatomy can be used to vary the volumes and shapes of some tissues. However, the  scaling of organs, even when different factors are used in orthogonal directions, does not fully and realistically capture the anatomical variations of organs within different human bodies. However, for many applications, having a population of phantoms that models the variations in patient anatomy and, in nuclear medicine, uptake realization is essential for comprehensive validation and training of image processing and reconstruction algorithms. To solve this, in \cite{Segars2013}, Segars et al. used a deformable image registration technique to map phantom labels to segmented patient images; the resulting deformation fields were then applied to the phantom, thus creating a population of new XCAT models that capture the anatomical variability among patients. This method relies on the segmentation of patient images, which is tedious and time consuming. In this work, we propose a Convolutional neural networks (ConvNets) based approach to perform phantom to patient registration. The resulting deformation field can then be applied to organ label maps to generate a gold-standard segmentation for the deformed phantom image.

\begin{figure*}[!t]
\centering
  \includegraphics[trim={3cm 5cm 7cm 0.5cm},clip, width=0.8\textwidth]{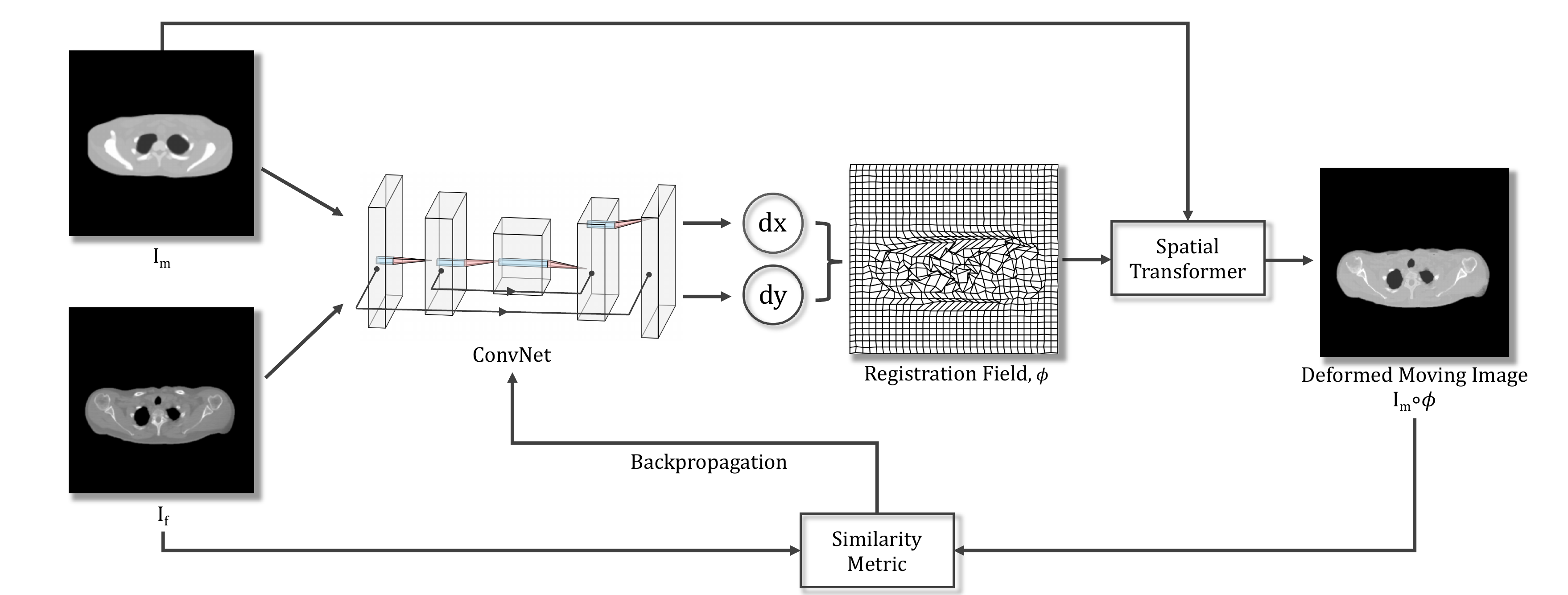}
  \caption{Schematic of the proposed method. The network takes a pair comprised of one moving and one fixed image as its inputs. The ConvNet learns from a single image pair and generates a deformation field, $\phi$. We then warp the moving image $I_m$ with $\phi$ using a B-spline spatial transformer. The loss determined by the image similarity measure between $I_m \circ \phi$ and $I_f$ is then backpropagated to update the parameters in the ConvNet. Since no aspect of the ConvNet is learned from a prior training stage, the method follows a fully unsupervised paradigm.}
  \label{fig_arch}
\end{figure*}

Deformable Image registration is a process of transforming two images into a single coordinate system, where one image is often referred to as the moving image, denoted by $I_m$, and the other is referred to as the fixed image, denoted by
$I_f$. Traditional methods formulate registration as a variational problem for estimating a smooth mapping, $\phi$, between the points in one image and those in another. They often tend to iteratively minimize the following energy function (eq. \ref{e1}) on a single image pair \cite{Sotiras2013}:

\begin{equation}
\label{e1}
    E=E_{sim}(I_m \circ \phi, I_f) + R(\phi),
\end{equation}
where, $E_{sim}$ measures the level of alignment between the transformed moving image, $I_m \circ \phi$, and the fixed image, $I_f$. Some common choices for $E_{sim}$ are mean squared error (MSE) or the squared $L^2$ norm of the difference \cite{Beg2005}, sum of squared differences (SSD) \cite{Wolberg2000}, cross-correlation (CC) \cite{Avants2008}, and mutual information (MI) \cite{Viola1997}. The transformation, $\phi$, at every point is defined by an identity transformation with the displacement field $\textbf{u}$, or $\phi = Id + \textbf{u}$, where $Id$ represents the identity transform \cite{Balakrishnan2019}. The second term, $R(\phi)$, is referred to as the regularization of the deformation, $\phi$, which enforces spatial smoothness. It is usually characterized by the gradients of $\textbf{u}$. One common assumption is that similar structures are present in both moving and fixed images. Hence, a continuous and invertible deformation field (a diffeomorphism) is desired, and the regularization term, $R(\phi)$, is designed to enforce or encourage this. Diffeomorphisms can be essential in some studies, for which the registration field is analyzed further. However, in the application of registration-based segmentation, the quality of the segmentation propagation is more critical than the diffeomorphic property of the underlying deformation fields \cite{Rueckert2006}. In this study, due to the large interior and exterior shape variations between digital phantoms and patients, diffeomorphism is less important. However, we show that by introducing various regularizers to the proposed model, the number of non-invertible voxel transformations in the resulting deformation field can be substantially reduced.

\begin{figure*}[!t]
\centering
  \includegraphics[trim={0cm 5.5cm 0cm 3.5cm},clip, width=0.8\textwidth]{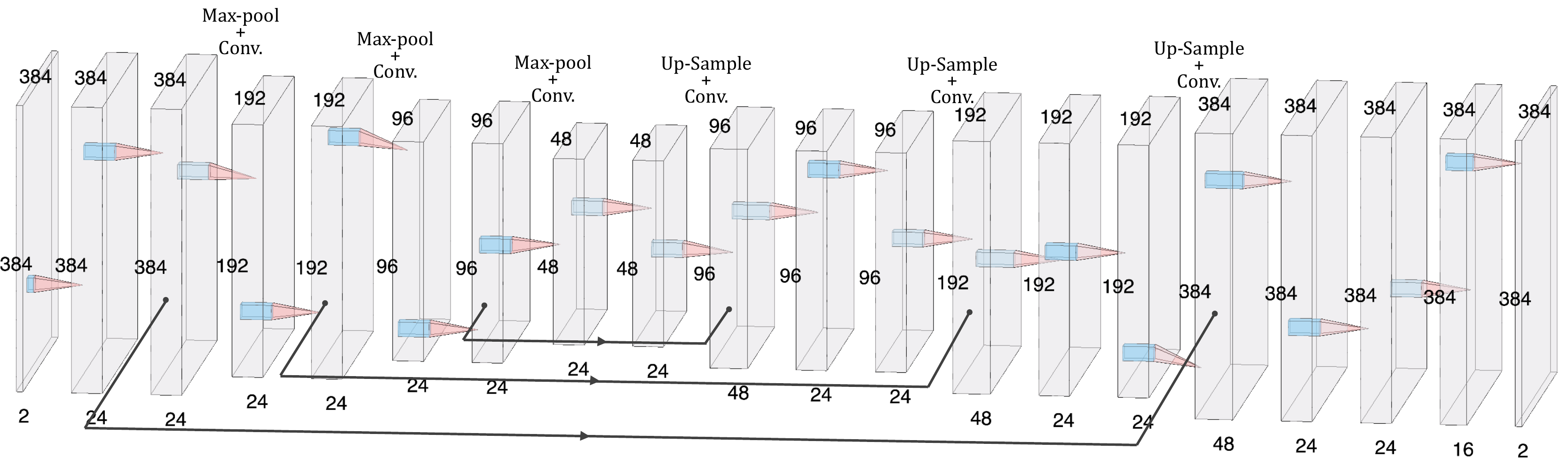}
  \caption{The ConvNet has a U-Net-like architecture.}
  \label{fig_unet}
\end{figure*}

Recently, many deep learning-based methods have been proposed to perform registration tasks, for instance \cite{Balakrishnan2019, Dalca2019, DeVos2017, Balakrishnan2018, Krebs2019}. Some of the listed methods were introduced as unsupervised (or more percisely, self-supervised) techniques, but they still require a prior training stage with a large amount of training data. These methods assume that neural networks can provide a universal and generalized model for image registration by minimizing the registration energy function over a dataset of images. This is a common assumption with deep-learning-based approaches. Yet, such an assumption could be unreliable according to a recent study from Zhang et al. \cite{Zhang2016}, where they showed that a well-generalized CNN classifier trained by a large dataset can still easily overfit a random labeling of the training data. Other studies on fooling deep neural networks (DNNs) with adversarial images also suggest that the well-trained networks can be unstable to small or even tiny perturbations of the data \cite{Su2019, Moosavi-Dezfooli2016, Goodfellow2014, Papernot2016, Szegedy2013}. On the other hand, the proposed registration method is fully unsupervised, meaning that \textbf{\textit{no previous training is required}}. Instead of following the conventional pattern of training a network on a large dataset of accurately annotated images, we show that a CNN can estimate an optimal deformation field for a single image pair by minimizing the energy function described in eq. \ref{e1} iteratively. This idea was inspired by Lempitsky et al.'s work on the Deep Image Prior\cite{Lempitsky2018} (DIP), where they showed that learning from a large amount of data is not necessary for building useful image priors, but the structure of a convolutional generator network itself is sufficient to capture image statistics. They treated the training of ConvNets with random initialization as a regularization prior, and in order to achieve good solutions in their application of image denoising, determining early stopping points was often required. Whereas in image registration, instead of starting from a random initialization (i.e., random noise images), it makes logical sense to initialize the ConvNet with a moving image. Since one would like to transform the moving image so that it is similar to a target image as possible, early stopping is not desired. In this work, we treat ConvNet as an optimization tool, where it minimizes the difference between moving and fixed images by updating its parameter values in each iteration.

\section{Method}
\subsection{Computerized Phantom Generation}
The phantom used in this study was created from the 3D attenuation distribution of the realistic NURBS-based XCAT phantom \cite{Segars2008}. Attenuation values were computed based on the material compositions of the materials and the attenuation coefficients of the constituents at 140 keV, the photon energy of Tc-99m. This single 3D phantom image was deformed to multiple patient CT images. The simulated attenuation map image can be treated as the template image, and phantom label map can then be thought of as the atlas in the traditional paradigm of medical image registration. The aim is to first register the phantom attenuation map image to patient CT images. Next, the registration parameters would be applied to the XCAT phantom label map (used to define organs and thus the activity distribution) to create new anthropomorphic phantoms. For the nuclear medicine imaging application, new images would be generated from the resulting phantoms using conventional physics-based simulation codes \cite{Frey1993, Kadrmas1996, Schuemann2014,Du1039547,Li2019}.
\subsection{Image Registration with ConvNet}
Let the moving image be $I_m$, and the fixed image be $I_f$; we assume that they are grayscale images defined over a n-dimensional spatial domain $\Omega \subset \mathcal{R}^n$ and affinely aligned. This paper primarily focuses on the 2D case, but the implementation is dimension independent. We model the computation of the displacement field, $\phi$, given the image pair, $I_m$ and $I_f$, using a deep ConvNet with parameters $\theta$, i.e., $f_\theta(I_m, I_f) = \phi$. Fig. \ref{fig_arch} describes the architecture of the proposed method; it consists of a ConvNet that outputs a registration field, and a B-spline spatial transformer. First, the ConvNet generates the $\phi$ for the given image pair, $I_m$ and $I_f$. Second, the deformed moving image is obtained by applying a B-spline spatial transformer that warps $I_m$ with $\phi$ (i.e., $I_m\circ\phi$). Finally, we backpropagate the loss computed from the similarity measure between $I_m\circ\phi$ and $I_f$ to update $\theta$ in the ConvNet. The steps are repeated iteratively until the loss converges; the resulting $\phi$ then represents the optimal registration field for the given image pair. The loss function ($\mathcal{L}$) of this problem can be formulated mathematically as:

\begin{equation} \label{e2}
\begin{split}
\mathcal{L}(I_m, I_f, \phi;\theta) &= \mathcal{L}_{sim}(I_m\circ \phi, I_f;\theta) + \lambda R(\phi;\theta)\\
&= \mathcal{L}_{sim}(I_m\circ f_\theta(I_m, I_f),I_f;\theta)\\
&\ \ \ \ \ \ \ + \lambda \mathcal{R}(f_\theta(I_m, I_f),\theta).
\end{split}
\end{equation}

\noindent where $\mathcal{L}_{sim}$ is the image similarity measure and $\mathcal{R}$ represents the regularization of $\phi$ Then, the parameters $\theta$ that generate the optimal registration field can be estimated by the minimizer:

\begin{equation}
\label{e3}
    \theta^*=\argmin_{\theta}\mathcal{L}(I_m, I_f, \phi;\theta),
\end{equation}
and the optimal $\phi$ is then given by:
\begin{equation}
\label{e4}
    \phi^*=f_{\theta^*}(I_m, I_f).
\end{equation}
Different choices of image similarity metrics and registration field regularizers ($R(\phi)$) were also studied in this work, and they are described in detail in a later section. The next subsection describes the design of ConvNet architecture.  
\subsubsection{ConvNet Architecture}
 The ConvNet had a U-Net-like "hourglass" architecture \cite{UNET}. The network consisted of one encoding path, which takes a single input formed by concatenating the moving and fixed images into a $2\times M\times M$ volume, where $M\times M$ represents the shape of one image. Each convolutional layer had a 3$\times$3 filter followed by a rectified linear unit (ReLU), and the downsampling was performed by 2$\times$2 max pooling operations. In the decoding stage, the upsampling was done by "up-convolution"\cite{UNET}. Each of the upsampled feature maps in the decoding stage was concatenated with the corresponding feature map from the encoding path. The output registration field, $\phi$, was generated by the application of sixteen 3$\times$3 convolutions followed by two 1$\times$1 convolutions to the 16 feature maps. This is a relatively small network with 98,794 trainable parameters in total. The network architecture is shown schematically in Fig. \ref{fig_unet}.

\subsubsection{Spatial Transformer}
The spatial transformer applies a non-linear warp to the moving image, where the warp is determined by a flow field of displacement vectors ($\textbf{u}$) that define the correspondences of pixel intensities in the output image to the pixel locations in the moving image. The intensity at the each pixel location, $\textbf{p}$, in the output image, $I_m\circ\phi(\textbf{p})$, is defined by:
\begin{equation}
\label{e:sp}
\begin{split}
    I_m\circ\phi(\textbf{p}) = I_m(\textbf{p}-\textbf{u}(\textbf{p})).
\end{split}
\end{equation}
Notice that $\textbf{p}-\textbf{u}(\textbf{p})$ is not necessarily integer, and pixel intensities are only defined at integer locations in the image. Therefore, the value of $I_m(\textbf{p}-\textbf{u}(\textbf{p}))$ was obtained by applying interpolation methods to the nearest pixels around $\textbf{p}-\textbf{u}(\textbf{p})$. We used, respectively, bi-linear and nearest-neighbor interpolation to obtain pixel values in the deformed XCAT phantom and the deformed labels (SPECT phantom).

\subsubsection{Image Similarity Metrics}
Over the years, considerable effort has been expended designing image similarity metrics. We mentioned some of the metrics that have been widely adopted in image registration in the previous section. In this work, we studied the effectiveness of five different loss functions, we also propose a new $\mathcal{L}_{sim}$ that combines the advantages of both Pearson's Correlation Coefficient (PCC) and the Structural Similarity Index (SSIM). In the following subsections, we denote the deformed moving image as $I_d$  (i.e., $I_d = I_m\circ\phi$) for simplicity.
\paragraph{Mean Squared Error (MSE)}
MSE is a measurement of fidelity, and indicates the degree of agreement of intensity values in images; it is applicable when $I_f$ and $I_m$ have similar contrast and intensity distributions. MSE is given by:

\begin{equation}
\label{e5}
    \text{MSE}(I_d,I_f)=\frac{1}{\Omega}\sum_{i\in\Omega}\Vert I_f(i)-I_d(i)\Vert^2,
\end{equation}
where $\Omega$ is the spatial domain. Then, the similarity loss function can be defined as $\mathcal{L}_{sim}(I_m, I_f, \phi;\theta)=\text{MSE}(I_d,I_f)$.

\paragraph{Pearson's Correlation Coefficient (PCC)}
PCC measures the linear correlation between two images. Unlike MSE, PCC is less sensitive to linear transformations of intensity values from one image to another. Its application to medical image registration is described in \cite{Saad2009}. PCC is defined as the covariance between images divided by the product of their standard deviations:
\begin{equation}
\label{e8}
\begin{split}
    &\text{PCC}(I_d,I_f)=\\ &\frac{\sum_{i\in\Omega}(I_f(i)-\bar{I}_f)(I_d(i)-\bar{I}_d)}{\sqrt{\sum_{i\in\Omega}(I_f(i)-\bar{I}_f)}\sqrt{\sum_{i\in\Omega}(I_d(i)-\bar{I}_d)}}
\end{split}
\end{equation}
where $\bar{I}_f$ and $\bar{I}_d$ represent the mean intensities. PCC has a range from -1 to 1, where 0 implies that there is no linear correlation, and -1 and 1 correspond, respectively, to the maximum negative and positive correlations between two images. Since a positive correlation is desired, we can define the loss function to be: $\mathcal{L}_{sim}(I_m, I_f, \phi;\theta)=1-\text{PCC}(I_m\circ\phi,I_f)$.

\paragraph{Local Cross Correlation (CC)}
\label{para:CC}
Another popular image similarity metric is CC, due to its robustness to intensity variations between images. It can be formulated as follows \cite{Balakrishnan2018, Avants2008, Zhu2019}:
\begin{equation}
\label{e6}
\begin{split}
    &\text{CC}(I_d,I_f)=\\
    &\sum_{p\in\Omega}\frac{(\sum_{p_i}(I_f(p_i)-\bar{I}_f(p))(I_d(p_i)-\bar{I}_d(p)))^2}{\sum_{p_i}(I_f(p_i)-\bar{I}_f(p))\sum_{p_i}(I_d(p_i)-\bar{I}_d(p))},
\end{split}
\end{equation}
where $p_i$ represents the pixel location within a window $p$, and $\bar{I}_f$ and $\bar{I}_d$ denote the local mean intensities within the window (we set $p$ to be a $3\times3$ window). Since $\text{CC} \geq 0$, we minimize the negative CC, the loss function is $\mathcal{L}_{sim}(I_m, I_f, \phi;\theta)=-\text{CC}(I_m\circ\phi,I_f)$.
\paragraph{Mutual Information (MI)}
MI was first applied to image registration in \cite{Maes1997}. It measures the statistical dependence between the intensities of corresponding pixels in both moving and fixed images. Let $p_{I_f}(a)$ and $p_{I_d}(b)$ be the marginal probability distributions of the fixed and deformed moving images. MI is a measure of the Kullback-Leibler Divergence \cite{Vajda1989} between the joint distribution $p_{I_fI_d}(a,b)$ and the distribution associated with the case
of complete independence $p_{I_f}(a)\cdot p_{I_d}(b)$\cite{Maes1997}:

\begin{equation}
\label{eMI}
\begin{split}
    \text{MI}(I_d,I_f)=\sum_{a,b}p_{I_fI_d}(a,b)\log\frac{p_{I_fI_d}(a,b)}{p_{I_f}(a)\cdot p_{I_d}(b)}.
\end{split}
\end{equation}
\noindent The joint distribution, $p_{I_fI_d}(a,b)$, can be computed as:
\begin{equation}
\label{eJD}
\begin{split}
    p_{I_fI_d}(a,b)=\frac{1}{\Omega}\sum_{i\in\Omega}\delta(I_f(i)-a)\delta(I_d(i)-b).
\end{split}
\end{equation}
\noindent Notice that the dirac-delta function is not differentiable, therefore the resulting loss cannot be back-propagated in the network. To solve this issue, we approximate $p_{I_fI_d}(a,b)$ with the differentiable Gaussian functions:
\begin{equation}
\label{eJD_approx}
\begin{split}
    p_{I_fI_d}(a,b)=\frac{1}{\Omega}\sum_{i\in\Omega}\frac{1}{\sigma^22\pi}e^\frac{(I_f(i)-a)^2}{2\sigma^2}e^\frac{(I_d(i)-a)^2}{2\sigma^2},
\end{split}
\end{equation}
where $\sigma$ is a user defined parameter that can vary depending on the images of a certain application. A larger $\sigma$ potentially leads to an improvement of computational efficiency, because the number of intensity bins used in the estimation of the marginal probability distributions could be reduced. Finally, the two marginal probability distributions can be derived using $p_{I_fI_d}(a,b)$. Maximizing MI is equivalent to minimizing the negative of the MI. Thus, the loss function is formulated as $\mathcal{L}_{sim}(I_m, I_f, \phi;\theta)=-\text{MI}(I_m\circ\phi,I_f)$.

\paragraph{Structural Similarity Index (SSIM)}
SSIM was proposed in \cite{Wang2004} for robust image quality assessments based on the degradation of structural information. Within a given image window, SSIM is defined by:
\begin{equation}
\label{e7}
\begin{split}
    &\text{SSIM}(I_d,I_f)=\frac{(2\mu_{I_d}\mu_{I_f}+C_1)(2\sigma_{I_fI_d}+C_2)}{(\mu_{I_f}^2+\mu_{I_d}^2+C_1)(\sigma_{I_f}^2+\sigma_{I_d}^2+C_2)},
\end{split}
\end{equation}
where $C_1$ and $C_2$ are small constants needed to avoid instability, $\mu_{I_f}$ and $\mu_{I_d}$, and $\sigma_{I_f}$ and $\sigma_{I_d}$ are local means and standard deviations of the images $I_f$ and $I_d$, respectively. The SSIM has a range from -1 to 1, where 1 indicates a perfect structural similarity. Thus, $\mathcal{L}_{sim}(I_m, I_f, \phi;\theta)=1-\text{SSIM}(I_m\circ\phi,I_f)$.

\begin{figure}[H]
\centering
\includegraphics[trim={5.5cm 6cm 6.5cm 5cm},clip,width=3in]{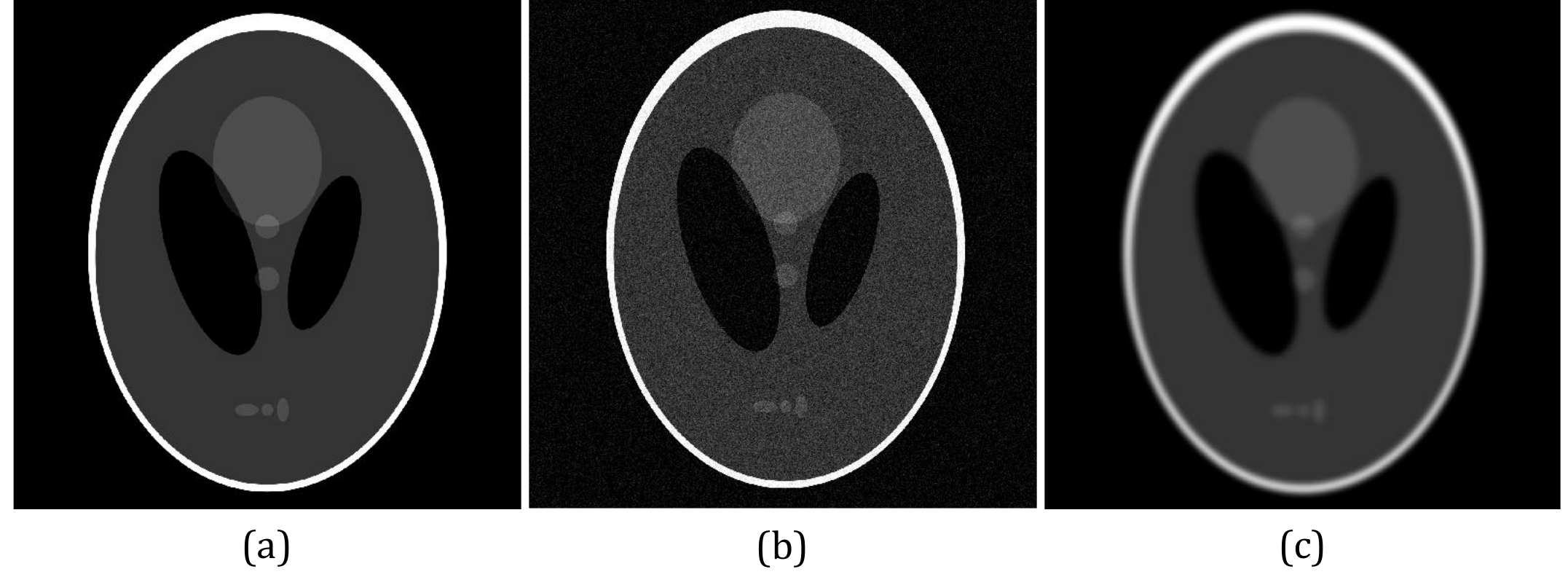}
\caption{Comparison of "Shepp-Logan" phantom images \cite{Shepp1974} with different types of distortions. (a) Original Image. (b) Image corrupted by Gaussian noise. SSIM: 0.14, PCC: 0.96. (c) Gaussian blurred image. SSIM: 0.9, PCC: 0.94.}
\label{fig_sim}
\end{figure}
\paragraph{PCC + SSIM}
While PCC is robust to noises, it was also found to be less sensitive to blurring. A motivating example is shown in Fig. \ref{fig_sim}, where in (b), the ”Shepp-Logan” phantom image \cite{Shepp1974} was corrupted with Gaussian noise, and in (c), the image was blurred by a Gaussian filter. Both (b) and (c) yield a lower SSIM and a higher PCC. If we think of (a) as a moving image, and (b) and (c) as fixed images, SSIM would impose the ConvNets to model the details, including noises and artifacts. Whereas, using PCC alone as the loss function might converge to a less accurate result. Hence, there is a need to balance the two similarity measures. Both PCC and SSIM are bounded with a range from -1 to 1, where 1 indicates the most similar. Thus, we propose to combine SSIM and PCC with an equal weight:
\begin{equation}
\label{e9}
\begin{split}
\mathcal{L}_{sim}(I_m, I_f, \phi;\theta)&=0.5*(1-\text{SSIM}(I_m\circ\phi,I_f))\\
&\ \ \ +0.5*(1-\text{PCC}(I_m\circ\phi,I_f))
\end{split}
\end{equation}


\subsection{Deformation Regulartization}
Optimizing the image similarity metrics would encourage the deformed moving image, $I_f$, to be as close as possible to the fixed image, $I_m$. However, the resulting deformation field might not be smooth or realistic. To impose smoothness and weakly enforce diffeomorphism in the deformation field, we tested several different regularizers.
\subsubsection{Diffusion Regularizer}
\label{sec_diff}
Balakrishnan, et al. used a diffusion regularizer in a ConvNet-based image registration model, VoxelMorph \cite{Balakrishnan2019}. In this method, the regularization is applied on the spatial gradients of the displacement field $\textbf{u}$:
\begin{equation}
\label{e10}
\begin{split}
\mathcal{R}_{Diffusion}(\phi;\theta)&=\sum_{i\in\Omega}\Vert\nabla\textbf{u}(i)\Vert^2
\end{split},
\end{equation}
where the spatial gradients are approximated by the forward difference, that is $\nabla\textbf{u}(i) \simeq \textbf{u}(i+1)-\textbf{u}(i)$. Minimizing this the value of this regularizer leads to smaller spatial variations in the displacements, resulting in a smooth deformation field.
\subsubsection{Total Variation Regularizer}
Instead of using the squared $L^2$ norm as the diffusion regularizer, the total variation norm regularizes the $L^1$ norm on the spatial gradients of $\textbf{u}$ \cite{HLi2018}:
\begin{equation}
\label{e11}
\begin{split}
\mathcal{R}_{TV}(\phi;\theta)&=\sum_{i\in\Omega}\Vert\nabla\textbf{u}(i)\Vert_1
\end{split}
\end{equation}
Penalizing the TV of the displacement field constrains its spatial incoherence without forcing it to be smooth. Detailed properties of TV regularization of displacements were studied by Vishnevskiy et al. in \cite{Vishnevskiy2017}.

\subsubsection{Non-negative Jacobian}
The determinants of the Jacobian represent the amount of transformation under a certain deformation. In \cite{Kuang2019}, Kuang et al. proposed a regularizer that specifically penalizes "folding" or non-invertable deformations, that is, the spatial locations where the Jacobian determinants are less than 0. This regularizer is formulated as:

\begin{equation}
\label{e12}
\begin{split}
\mathcal{R}_{Jacobian}(\phi;\theta)&=\sum_{i\in\Omega}(\vert det(\textbf{J}_\phi(i))\vert-det(\textbf{J}_\phi(i)))
\end{split}.
\end{equation}

Combined this with diffusion regularization to constrain the overall smoothness results in a regularizer that produces deformations with fewer folded pixels:
\begin{equation}
\label{e13}
\begin{split}
\mathcal{R}_{reg}(\phi;\theta)=\mathcal{R}_{Diffusion}(\phi;\theta)+\alpha\mathcal{R}_{Jacobian}(\phi;\theta),
\end{split}
\end{equation}
where $\alpha$ is a weighting parameter.

\subsubsection{Gaussian Smoothing}
A direct way to constrain a deformation field to be smooth is to convolve the displacements with a Gaussian smoothing filter parameterized by its standard deviation, $\sigma$ \cite{Krebs2019}:
\begin{equation}
\label{e14}
\begin{split}
\hat{\textbf{u}}=G_{\sigma}\ast\textbf{u},
\end{split}
\end{equation}
where a larger $\sigma$ gives a smoother deformation, and vice versa.

\subsection{Registration Procedure}
The overall algorithm for the proposed method is shown in Algorithm. \ref{alg:ConvReg}. In the beginning, we initialized an untrained ConvNet ($f_{\theta}$) for a given pair of moving and fixed images, $I_m$ and $I_f$. First, the untrained $f_{\theta}$ produces an initial deformation field, $\phi$. Second, we deform the moving image with $\phi$ (i.e., $I_m\circ\phi$). Then, the registration loss is computed as:
\begin{equation}
\label{e15}
\begin{split}
\ell=\mathcal{L}_{sim}(I_d,I_f;\theta)+\lambda\mathcal{R}(\phi;\theta),
\end{split}
\end{equation}
where $\mathcal{L}_{sim}$ represents the similarity measure between $I_d$ and $I_f$, $\mathcal{R}$ represents the value of the regularizer applied to the deformation field, and $\lambda$ is a user-defined weighting parameter to control the effectiveness of $\mathcal{R}$. The loss is back-propagated to update the parameters in $f_{\theta}$. The above procedure is repeated for a pre-specified number of iterations. 
\begin{algorithm}
\caption{ConvNet Registration}\label{alg:ConvReg}
\begin{algorithmic}[1]
\Procedure{CNNReg}{$I_m,I_f$}\Comment{Input $I_m$ and $I_f$}
\State $f_{\theta}$ = \textit{\text{Initialize}}(\textit{\text{ConvNet}})
\While{$i<iter$}\Comment{For $iter$ number of iterations}
\State $\phi = f_{\theta}([I_m,I_f])$ \Comment{Predict deformation, $\phi$}
\State $I_d = I_m\circ\phi$ \Comment{Deform moving image, $I_m$}
\State $\ell=\mathcal{L}(I_d,I_f;\theta)+\lambda\mathcal{R}(\phi;\theta)$ \Comment{Compute loss}
\State $f_{\theta} = \textit{\text{BackPropagate}}(f_{\theta}, \ell)$ \Comment{Update ConvNet}
\State $i=i+1$
\EndWhile\label{euclidendwhile}
\State \textbf{return} $I_d, \phi$
\EndProcedure
\end{algorithmic}
\end{algorithm}

Since no information other than the given image pair is needed, the proposed method requires no prior training and is thus fully and truly unsupervised. The ConvNet is capable of learning an "optimal" deformation from a single pair of images. In the next section, we discuss a series of experiments that were performed to study the effectiveness of the proposed method.

\section{EXPERIMENTS}
\label{sec:exp}
The goal of this work was to create anthropomorphic phantoms by registering the XCAT phantom attenuation map with patient CT images, and then using the warped map of XCAT phantom labels map. Nine clinical low-dose whole-body CT patient scans were used in this study; for those, only the torso part of the scans was extracted, resulting in 1153 2D-transaxial slices in total. The proposed method was implemented using Keras with a Tensorflow \cite{tensorflow2015} backend on an NIVIDIA Quadro P5000 GPU. The patient CT data was obtained from a publicly available dataset (NaF Prostate, \cite{Kurdziel2012}) from The Cancer Imaging Archive (TCIA, \cite{Clark2013}). We first compared the performance produced by the ConvNets with different image similarity metrics. Then, we compared the proposed method to state-of-the-art registration algorithms: the symmetric image normalization method (SyN) \cite{Avants2008} from the ANTs package \cite{Avants2009}, and a learning-based self-supervised method, VoxelMorph \cite{Balakrishnan2018, Balakrishnan2019}. Finally, we applied the proposed method and show that the resulting registered image can be used in creation of realistic simulated single-photon emission computerized tomography (SPECT) images.

\begin{figure*}[!t]
\centering
  \includegraphics[ width=0.95\textwidth]{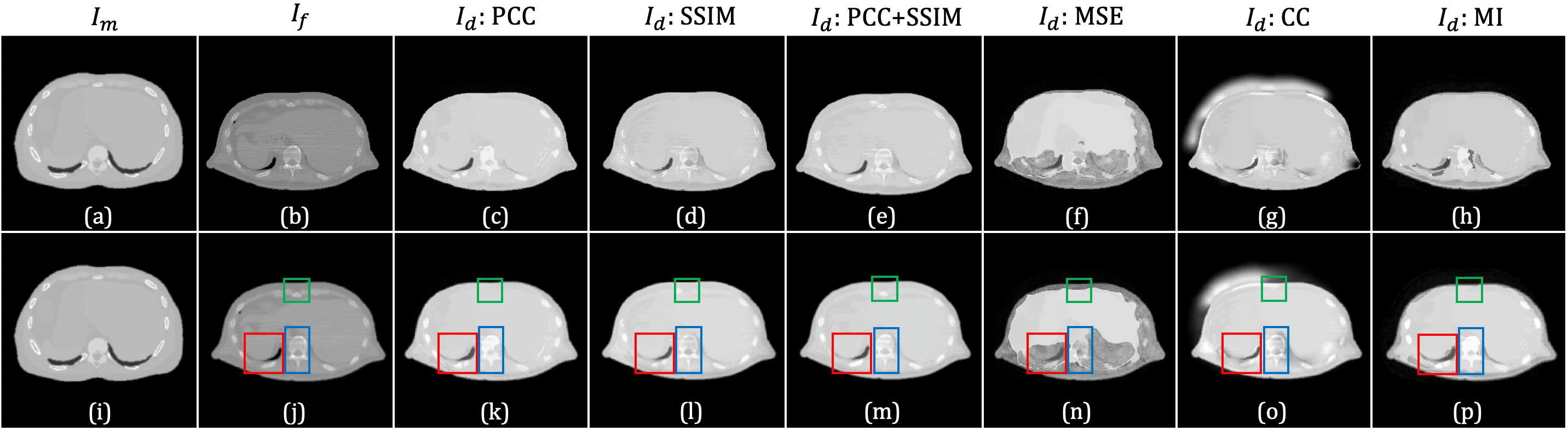}
  \caption{Comparison of registered XCAT phantom image generated using different loss functions (without any regularization), some differences are highlighted by colored rectangles. Top row exhibits the results generated without pre-filtering the fixed image; the bottom row shows the results generated using the pre-filtered fixed image. The images in the first two columns are: (a) and (i) Same example slice of attenuation map generated from the XCAT phantom, which served as the moving image, $I_m$; (b) and (j) are the same patient CT images, prior to use in the registration, image (j) is (i) blurred with Gaussian filter ($\sigma=0.8$) to reduce noise and artifacts. The images in (i) and (j) were used as the fixed image, $I_f$. Starting with column 3, images are shown resulted from applying the ConvNet using 3 different loss functions: (c) and (k) PCC; (d) and (l) SSIM; (e) and (m) PCC+SSIM; (f) and (n) MSE; (g) and (o) CC; and (h) and (p) MI.}
  \label{fig_imgs}
\end{figure*}

\begin{figure*}[!t]
\centering
  \includegraphics[trim={0cm 1cm 0cm 1cm},clip,width=6in]{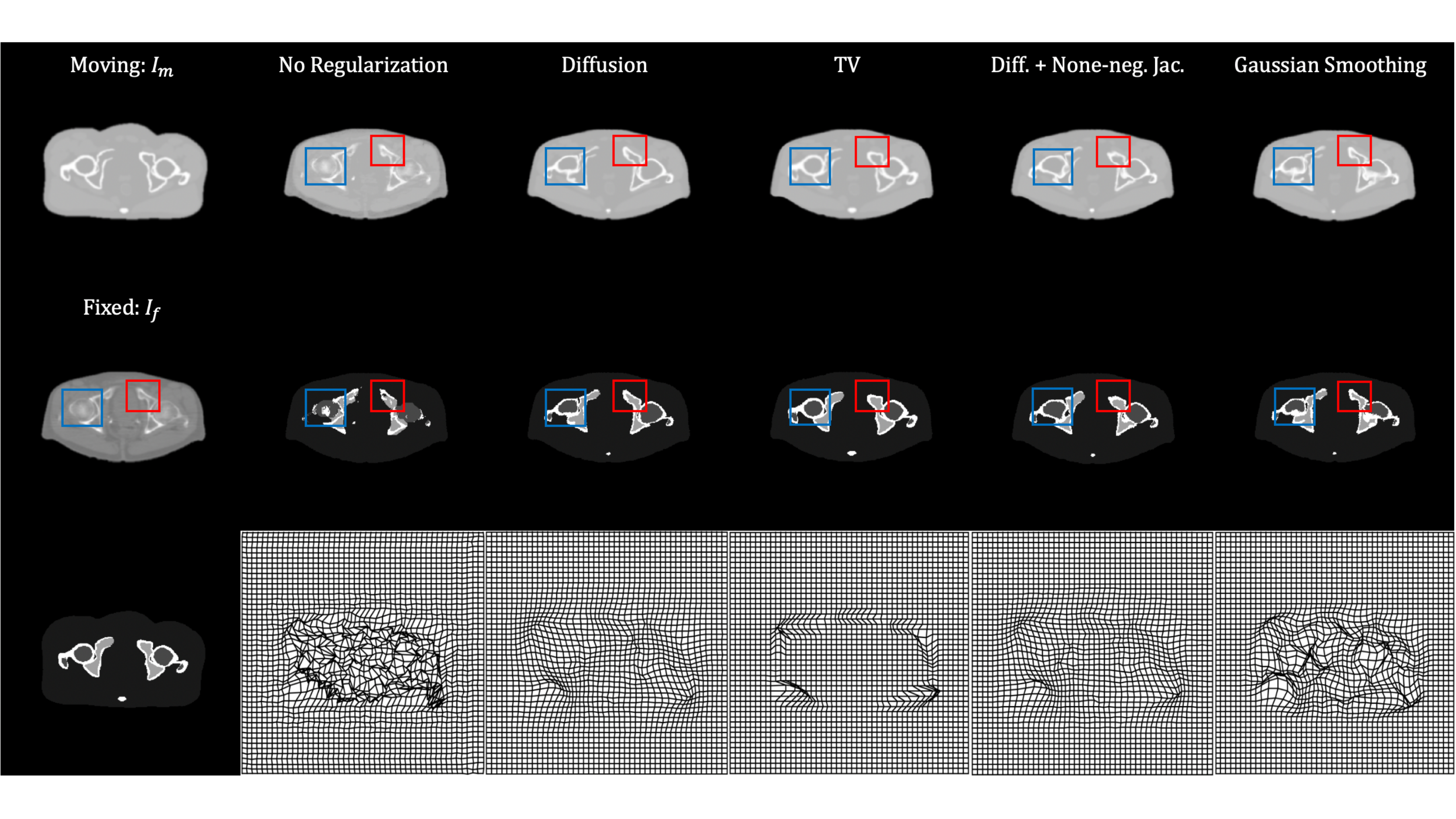} 
  \caption{Example results from different regularization techniques. The three figures in the first column represent moving image, fixed image, and the corresponding SPECT phantom, respectively. The second to last column shows deformed images: the first row shows the deformed XCAT phantom, the second row shows the deformed SPECT phantom, and the last row shows the deformed grid.}
  \label{fig_reg_img}
\end{figure*}

\begin{figure*}[!t]
\centering
  \includegraphics[trim={4cm 5cm 4cm 5cm},clip,width=5.3in]{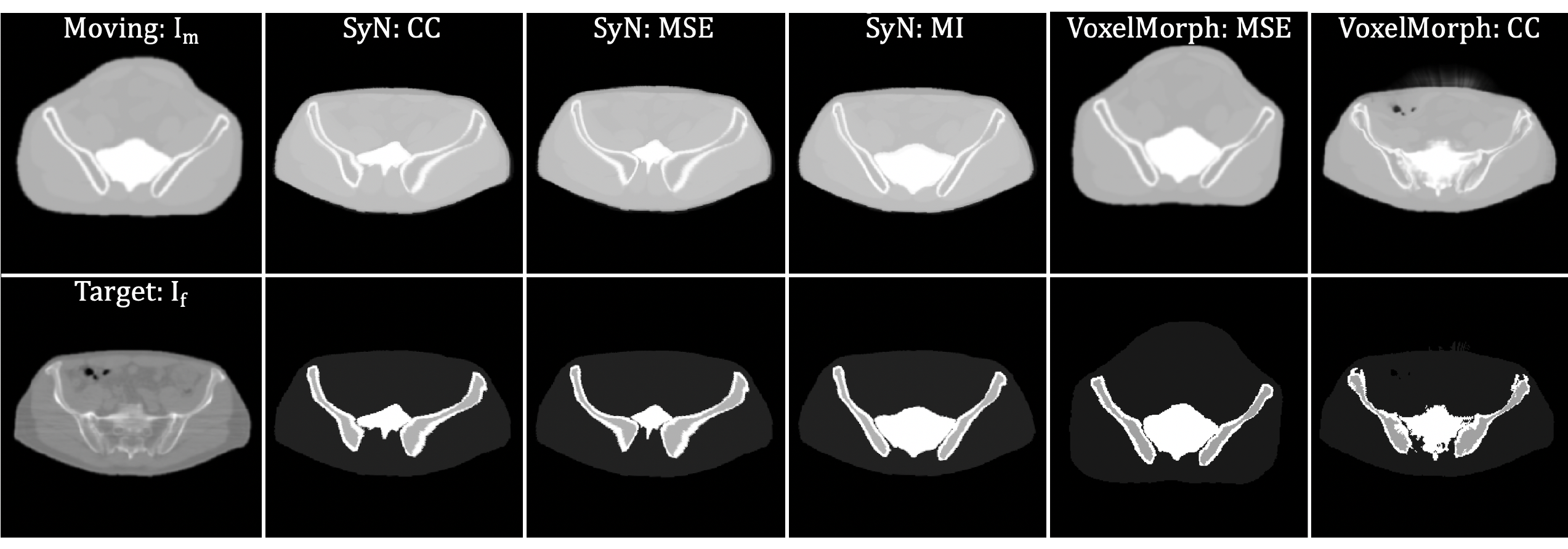} 
  \caption{Example results generated by two baseline methods, SyN and VoxelMorph. The \nth{1} column: moving (XCAT phantom) and target image (patient CT). For the second to the last column, the first row corresponds to the deformed moving images, and the second row shows the deformed label map (used to make the SPECT phantom). The \nth{2} through \nth{4} columns show results from SyN using CC, MSE, and MI. The \nth{5} column through the last column shows results from VoxelMorph using MSE and CC.}
  \label{fig_def_imgs_base}
\end{figure*}

\begin{figure*}[!t]
\centering
  \includegraphics[trim={4cm 5cm 4cm 5cm},clip,width=5.3in]{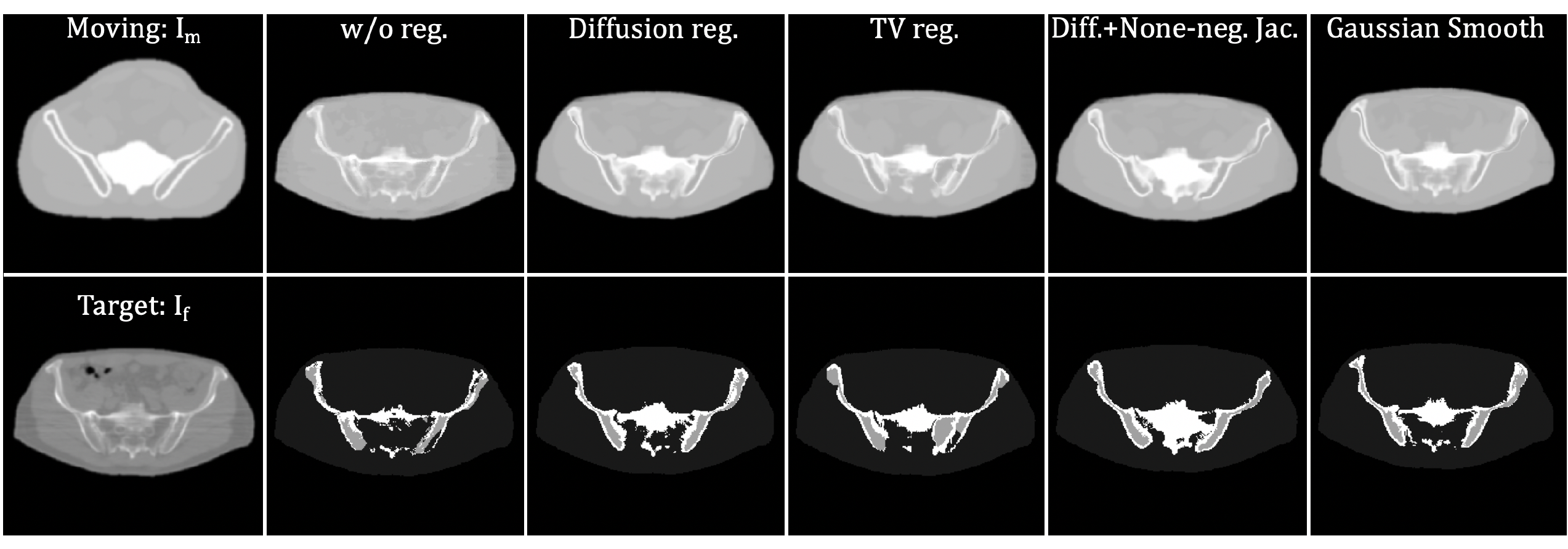} 
  \caption{Example results generated by the proposed method with different regularization techniques. The \nth{1} column exhibits moving (XCAT phantom) and target image (patient CT). The \nth{2} column to the last column display the deformed results. These correspond, respectively, to no regularization, diffusion regularization, TV norm, diffusion with none-negative Jacobian regularization, and Gaussian smoothing.}
  \label{fig_def_imgs_ours}
\end{figure*}

\begin{table*}[h]
\captionsetup{justification=centering}
  \centering
  \caption{Quantitative Comparisons Different Registration Methods}
  \begin{tabular}{ p{5.5cm}|p{2.2cm}|p{2.2cm}|p{2.2cm}|p{2.2cm}  }
 
 Method        & SSIM & MSE & $\vert\textbf{J}_{\phi}\vert \le 0$ (counts) & $\%$ of $\vert\textbf{J}_{\phi}\vert \le 0$ (\%)\\
 \hline
 \hline
 Affine only& $0.828\pm0.008$ & $69.213\pm2.748$ & - & -\\
 \hline
 VoxelMorph (MSE) \cite{Balakrishnan2019} & $0.878\pm0.003$ & $47.008\pm2.436$ & $685\pm185$ & $0.5\pm0.1$ \\
  \hline
  VoxelMorph (CC) \cite{Balakrishnan2019} & $0.916\pm0.006$ & $43.457\pm4.829$ & $2754\pm370$ & $1.9\pm0.3$ \\
  \hline
 SyN (MSE) \cite{Avants2008}     & $0.884\pm0.011$ & $51.999\pm4.135$ & - & -\\
 \hline
 SyN (MI) \cite{Avants2008}  &  $0.881\pm0.011$   & $55.059\pm3.996$ & - & -\\
 \hline
 SyN (CC) \cite{Avants2008} & $0.886\pm0.011$ & $52.838\pm4.138$ & - & -\\
 \specialrule{.1em}{.05em}{.05em} 
 UnsupConvNet (w/o regularization) & $\textbf{0.955}\pm0.007$ & $\textbf{37.340}\pm5.078$ & $21082\pm3938$ & $14.3\pm2.7$ \\
 \hline
 UnsupConvNet (w/ diffusion regularization) & $\textit{0.930}\pm0.008$ & $\textit{42.602}\pm5.371$ & $1202\pm225$ & $0.8\pm0.1$ \\
  \hline
  UnsupConvNet (w/ diff. + None-neg. Jac. reg.) & $0.915\pm0.009$ & $44.788\pm4.856$ & $518\pm74$ & $0.4\pm0.1$ \\
 \hline
 UnsupConvNet (w/ TV regularization) & $0.870\pm0.030$ & $54.691\pm9.311$ & $659\pm459$ & $0.4\pm0.3$ \\
 \hline
 UnsupConvNet (w/ Gaussian filtering) & $\underline{0.939}\pm0.008$ & $\underline{41.502}\pm5.419$ & $8500\pm1829$ & $5.7\pm1.3$ \\
\end{tabular}
  \caption*{Comparison of SSIM, MSE, and the number and percentage of pixel locations with non-positive Jacobian determinant among the proposed method (UnsupConvNet), SyN, and VoxelMorph. The top three results in SSIM and MSE are shown in \textbf{bold}, \underline{underline}, and \textit{italics}, respectively. Evaluations were done on 2D images with size $384\times384$.}
  \label{tab:1}
\end{table*}
\subsection{Loss Function Comparisons}
Some examples of the registered XCAT phantom images resulting from the six loss functions are shown in Fig. \ref{fig_imgs}. Images (a) and (h) represent the same moving image, and (b) and (i) are the target images from the same CT slice, where the later was blurred by a low-pass Gaussian filter to reduce the effects streaking artifacts. The third through last columns show results form regularizing using, left to right, PCC, SSIM, PCC+SSIM, MSE, CC, and MI. The MSE is a widely used loss function in both traditional and learning-based image registration methods, but it did not provide good results in this application (as visible from the sixth column in Fig. \ref{fig_imgs}). This is likely because MSE does not incorporate spatial information, and it is sensitive to linear transformations of the mean intensity values. CC is another commonly used metric that is seen in various applications, but it produced a suboptimal result with image artifacts (as shown in the seventh column in Fig. \ref{fig_imgs}). While PCC loss was robust to image artifacts, it produced "cartoonish" results around the spine (see the regions highlighted in rectangles in (c) and (j)). On the other hand, the SSIM loss function produced an image reproduced the noise and artifacts in the target image (as shown in (d)). The results produced by SSIM+PCC exhibits fewer image artifacts and provided the best structural match to the target image (as shown in (m)). Combined with the Gaussian pre-filtering to suppress streaking artifacts in the target image, SSIM+PCC generated the best qualitative results among the loss functions evaluated.

\subsection{Regularization Comparisons}
Fig. \ref{fig_reg_img} shows some results generated using different regularizers. The three images in the first column are a slice of XCAT phantom (moving image), a slice of patient CT image (fixed image), and a slice of SPECT phantom (i.e., the label map for the moving image). The second through last columns show the deformed moving image (first row), transformed labels (second row), and deformed grids (last row) by using no regularization, diffusion regularization, TV regularization, diffusion with none-negative Jacobian regularization, and Gaussian smoothing, respectively. The deformed moving image using no regularization has a virtually identical appearance compared to the fixed image. However, the deformed label maps were unrealistic: in the regions highlighted in rectangles, the bone marrow appears outside of the cortical bone. Applying regularization to the deformation field helped with this issue, but there was a clear trade-off between the similarity to the fixed image and the smoothness of the deformation field. This trade-off was quantitatively studied, and the results are discussed in the next section. We specifically quantify the regularity of the field by counting all the pixel deformations for which the transformation is not diffeomorphic (i.e., folding or $\vert\textbf{J}_{\phi}\vert \le 0$) \cite{Balakrishnan2019, ASHBURNER200795}.

\begin{figure*}[!t]
\centering
  \includegraphics[trim={4.5cm 3.8cm 3.5cm 3.5cm},clip,width=0.95\textwidth]{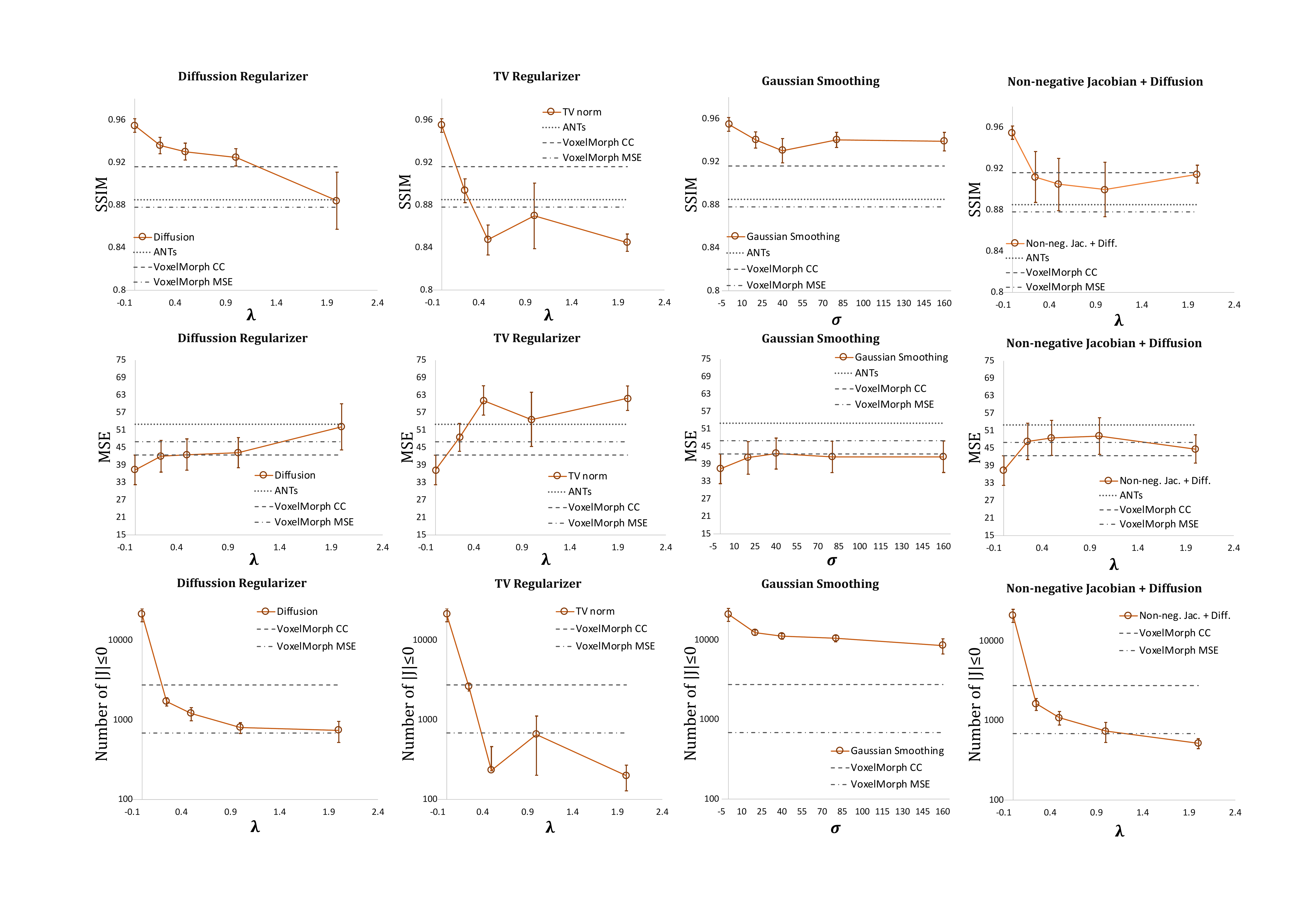} 
  \caption{The effect of different regularization parameters on the registration accuracy (SSIM and MSE), and deformation regularity (number of folded pixels, i.e. where the determinant of Jacobian was $\leq 0$). First to last rows indicate the performances in SSIM, MSE, and number of folded pixels, respectively. The columns, from left to right, are the results generated using the diffusion regularizer, TV regularizer, Gaussian smoothing, and non-negative Jacobian + diffusion regularizer, respectively.}
  \label{fig_eval_par}
\end{figure*}
\begin{figure*}[ht]
\centering
\includegraphics[width=1\textwidth]{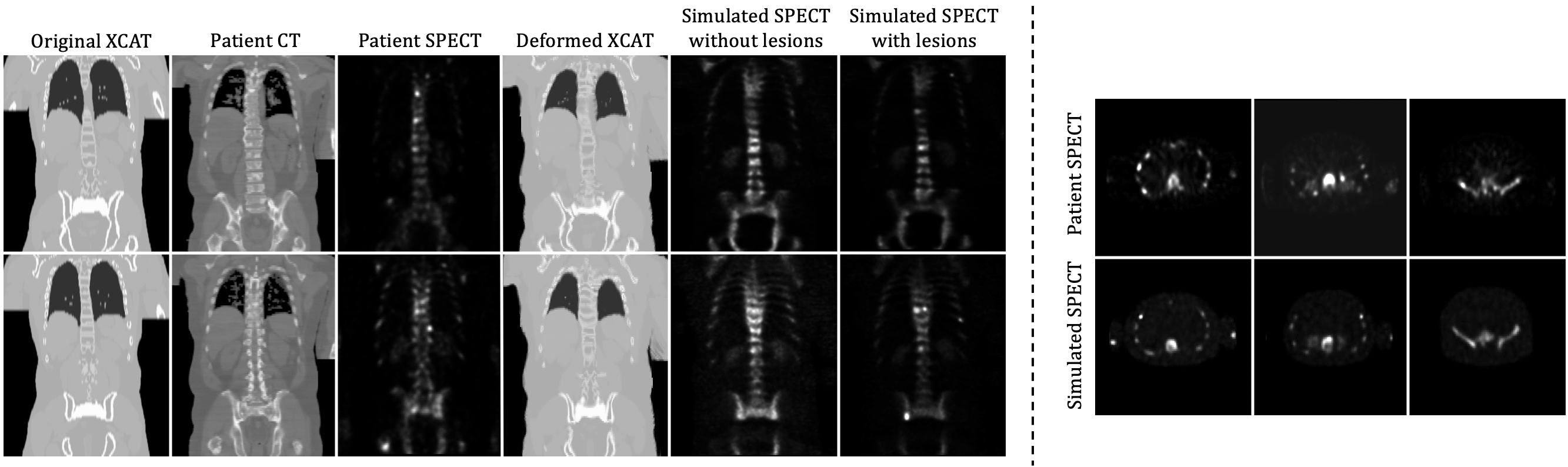}
\caption{Visualization of the deformed XCAT phantom and SPECT simulations. Left: Two coronal slices of the original XCAT, patient CT and SPECT scans, the registered XCAT, the SPECT simulation, and the simulated SPECT with lesion added. Right: Comparison of transverse slices between patient SPECT scan and SPECT simulation.}
\label{fig_simulation}
\end{figure*}
\subsection{Registration Performance Comparisons}
In this subsection, we compare the proposed method with the SyN \cite{Avants2008} and VoxelMorph \cite{Balakrishnan2018, Balakrishnan2019} algorithms. Since VoxelMorph requires prior training, it was evaluated using a leave-one-out method: images from eight patients were used for training (~1024 images), and images from one patient were treated as the test set (~128 images). Then, we altered patients whose image was used for  training and testing so that the evaluation 9 possible combinations of the patient images. Figs. \ref{fig_def_imgs_base} and \ref{fig_def_imgs_ours} show comparisons of the proposed method with different regularization, the SyN, and the VoxelMorph methods. The first column shows the moving and fixed images. The second to the last column shows the deformed XCAT images (upper row) and deformed labels (lower row). Based on these qualitative results, the proposed method provides a more detailed deformation than other methods, where the anatomy of bone structures and soft tissues were modeled precisely. Since a gold-standard bone segmentation is not available for the NaF Prostate dataset \cite{Kurdziel2012}, the registration performance was evaluated quantitatively based on MSE and SSIM between $I_m\circ\phi$ and $I_f$. The results are shown in Table. \ref{tab:1}. Without any regularization of the deformation field, the proposed method gives a mean SSIM of 0.955 and a mean MSE of 37.340, which outperforms the SyN and VoxelMorph by a significant margin (with $p\text{-values}<0.0001$ from the paired t-test). 

The plots in Fig. \ref{fig_eval_par} exhibit the impact of different regularization parameters on the SSIM, MSE, and the number of folded pixels. A decreasing trend in registration accuracy and the number of folded pixels was generally seen with increasing weighting parameter values ($\lambda$ and $\sigma$) for regularizers. Among the different regularization methods, the diffusion regularizer (column 1 in Fig. \ref{fig_eval_par}) with $\lambda=1$ yielded the best balance between registration accuracy and the number of folded pixels. Overall, the method achieved comparable performances to VoxelMorph in terms of deformation regularity (as measured by the number of pixels where there was folding) while providing better registration accuracy. 

\subsection{SPECT simulations}
In this section, we demonstrate a 3D application of the proposed method to generate realistic medical image simulations. We employed the proposed registration method to map the XCAT phantom to a patient CT scan acquired as part of a SPECT/CT acquisition. We then generated a realistic simulated SPECT image on the basis of the resulting deformed XCAT phantom as described below. The patient scan was acquired using a clinical whole-body SPECT/CT scan protocol; the CT scan was a low-dose one designed to provide an attenuation map. Both SPECT and CT images were reconstructed using scanner software. Two sample coronal slices of the patient scans are shown in the second and the third columns of the left panel in Fig. \ref{fig_simulation}; the second and third columns show the CT and SPECT images, respectively. We used the proposed method with a diffusion regularizer ($\lambda=1$) to perform the 3-dimensional registration. The resulting deformed XCAT phantom and the corresponding SPECT simulation are shown in the fourth and the fifth columns, respectively. We then added several artificial lesions to the phantoms, and two example slices of the resulting SPECT simulation are shown in the last column. SPECT projections were simulated by an analytic projection algorithm that realistically models attenuation, scatter, and the spatially-varying collimator-detector response \cite{Frey1993, Kadrmas1996}. We computed attenuation values on the basis of the material compositions and the attenuation coefficients of the constituents at 140 keV, the photon energy of Technetium-99m. We inserted several artificial sclerotic bone lesions to random bone regions with increased attenuation coefficient and radio-pharmaceutical uptake. The bones had an uptake of 12.6 times that of the soft-tissue background, and bone lesions had an uptake of 3.5-4.5 times that of normal bone. These scale factors were computed based on the patient SPECT scan. SPECT simulations were reconstructed using a the ordered subsets-expectation maximization algorithm (OS-EM) \cite{Hudson1994} \cite{He2005} using 5 iterations and 10 subsets. The figure on the right panel in Fig. \ref{fig_simulation} shows three transverse slices of the patient SPECT (top row) and the simulated SPECT (bottom row). Because the deformed XCAT phantom was able to successfully capture the anatomical structures in the patient scan, when combined with realistic physics models of image formation processing, the resulting SPECT simulation appears quite realistic compared to the patient SPECT scan. In addition, the relationship between the generated phantom activity distribution and the projection data is quantitatively realistic because of the method used to generate the projections.
\section{Conclusion}
We have developed a method to create anthropomorphic phantoms using an unsupervised, ConvNet-based, end-to-end registration technique. Unlike  existing ConvNet-based registration methods, the proposed method requires no prior training. While classical registration methods also do not require training data, they work in a lower-dimensional parameter space; the proposed approach operates directly in the high-dimensional parameter space common to deep-learning-based techniques without any prior training. Compared to the commonly used loss functions in ConvNet-based registration, we demonstrated that the registration performance can be improved by use of the combination of SSIM and PCC as a loss function for updating the parameters in the ConvNet. The proposed method was evaluated for the application of registering the XCAT phantom with real patient CT scans as part of a process to simulate realistic nuclear medicine images. We compared the registration performance of the proposed technique in terms of SSIM and MSE to conventional state-of-the-art image registration methods. Both quantitative and qualitative analyses indicated that the proposed method provided the best registration results. We also demonstrated that the proposed method, combined with accurate simulation tools, provided a highly realistic anthropomorphic medical image with known truth that faithfully represents the image formation process and qualitatively matches the appearance of a real patient image.

\section*{Acknowledgment}
This work was supported by a grant from the National Cancer Institute, U01-CA140204. The views expressed in written  materials or publications and by speakers and moderators do not necessarily reflect the official policies of the NIH; nor does mention by trade names, commercial practices, or organizations imply endorsement by the U.S. Government.

We would like to show our gratitude to Dr. Daniel Tward and Shuwen Wei for sharing  pearls of wisdom with us during the course of this research.

\bibliographystyle{IEEEtran}
\bibliography{ref}

\end{document}